\documentclass[prd,12pt]{revtex4}
\usepackage{amsmath,amssymb} 
\usepackage[dvips]{graphicx} 
\usepackage{graphics}
\usepackage{graphicx}
\usepackage{epsfig}
\usepackage[english]{babel}

\newcommand\nn{\nonumber}
\newcommand\ba{\begin{eqnarray}}
\newcommand\ea{\end{eqnarray}}

\begin{document}

\title{Contributions of semi-hadronic states $P\gamma;S\gamma, \pi^+\pi^-\gamma$ to \\ amm of muon,
in frames of Nambu-Jona-Lasinio model.}
\author{A.~I.~Ahmadov$^{a,b}$ \footnote{E-mail: ahmadov@theor.jinr.ru},
E.~A.~Kuraev$^a$ \footnote{E-mail: kuraev@theor.jinr.ru} and
M.~K.~Volkov$^a$ \footnote{E-mail: volkov@theor.jinr.ru}}
\affiliation{$^{a}$ JINR-BLTP, 141980 Dubna, Moscow region,
Russian Federation}
\affiliation{$^{b}$ Institute of Physics, Azerbaijan
National Academy of Sciences, Baku, Azerbaijan}

\begin{abstract}
We calculate the contribution of semi-hadronic states with pseudoscalar $P=\pi^0, \eta$
and scalar ($\sigma$(550))
meson accompanied with real photon as an intermediate state of a heavy photon
to the anomalous magnetic moment of muon.
We consider the intermediate states with $\pi_0$ and $\sigma$ as a hadrons
in frames of Nambu-Jona-Lasinio model. The contribution of $\pi_0\gamma$ state
is in agreement with results obtained in previous theoretical considerations as
well as with experimental data $a_\mu^{\pi_0\gamma}\approx 4.5 \times 10^{-10}$,
besides we estimate $a_{\mu}^{\eta\gamma}=0.7 \times 10^{-10},\,\,
a_{\mu}^{\sigma\gamma} \sim 1.5 \times 10^{-11}, \,\,
a_{\mu}^{\pi^+\pi^-\gamma} \sim 3.2 \times 10^{-10}.$
We discass as well the LbL mechanism with $a_{\mu}^{lbl}=10.5 \cdot 10^{-10}.$
\end{abstract}

\maketitle

\section{Introduction}
\label{Introduction}
One of modern precise test of Standard Model is the measurement of anomalous
magnetic moment of muon (amm) $a_{\mu}$. The SM contributions are usually split into
three parts:

$$a_{\mu}=a_{\mu}^{QED}+a_{\mu}^{EW}+a_{\mu}^{hadr}.$$

Contributions of hadrons associated with real photons (semi-hadrons ones) can
be separated to two classes. One of them consist in diagrams of vertex type
with heavy photon with insertion of hadronic vacuum polarization block (see Fig1).
Another contained the light-by-light scattering block (LbL) will be discussed  below (Fig.2).
Using the dispersion approach first type of contributions can be written as
\ba
a_{\mu}^{P(S)
\gamma}=\frac{1}{4\pi^3}\int\limits_{m_{P(S)}}^{\infty}ds \cdot
\sigma^{e \bar e \to P(S)\gamma}(s) \cdot
K\left(\frac{s}{M_{\mu}^2}\right).
\ea
The the analytic  form of the kernel $K(\rho)$ \cite{Brodsky} is:
\ba
K(\rho)=\int\limits_0^1\frac{x^2(1-x) d x}{x^2+(1-x)\rho}; \nn \\
K(\rho)=\frac{1}{2}-\rho+\frac{1}{2}\rho(\rho-2)\ln\rho-\frac{(\rho^2-4\rho+2)\rho}{2\sqrt{\rho(\rho-4)}}\nn \\
\ln\frac{\sqrt{\rho}+\sqrt{\rho-4}}{\sqrt{\rho}-\sqrt{\rho-4}};
\nn \\
\rho=\frac{s}{M_{\mu}^2};\,\,K^{(1)}(\rho)|_{\rho \gg 1}=\frac{1}{3\rho}.
\ea
The main part of
contribution to $a_{\mu}^{hadr}$ of order $5004$ (units $10^{-11}$ implied) arise
from $\pi^+\pi^-$ channel annihilation of $e^+e^-$ pair (3 $\pi$: 438 $\cdot 10^{-11}$;
2K: 314 $\cdot 10^{-11}$ \cite{Yndur02}).

Below we will consider the annihilation channels
\ba
e^+e^- \to \gamma^* \to P\gamma, \,\,S\gamma; \,\,\, P=\pi_0,\eta; \,\,S=\sigma, \nn \\
e^+e^- \to \gamma^* \to \pi^+\pi^-\gamma.
\ea

During the recent years the papers with calculation of semi-hadron states
where published \cite{Achasov02,Naris, Yndur02}. Rather stable results was
obtained for the $\pi_0\gamma$ state, whereas a contradictive results was obtained
for contribution of $\sigma\gamma$ state \cite{Naris}.
Below we obtain these contributions in frames NJL model \cite{Echaya,Prog,UFN}, both are consistent with
modern experimental data \cite{Akhmet}.

The relevant part of chiral Lagrangian in U(3)$\times$ U(3) chiral  NJL is \cite{IMP,CEPhys,Volkov09,Echaya}
\ba
L=\bar {q}[i \hat {\partial}+m-e Q \hat {A} +g_{\pi}(\lambda_3 \pi_0 +
\lambda_+ \pi_+ +\lambda_- \pi_-)\gamma_5 +g_{\sigma} \sigma \cdot \lambda_3 + \nn \\
g_k(\lambda_+ K_+ +\lambda_- K_-) +
\frac{g_{\rho}}{2}(\lambda_3 \hat {\rho_0}+
\lambda_4 \hat {\omega})]q,
\ea
where $\sigma=\lambda_u \sigma_u +\sigma_s \lambda_s$, $\bar q =(\bar {u}, \bar {d}, \bar {s})$ where $u$, $d$, $s$ are the quark fields, Q=diag(2/3,-1/3, -1/3)
is, the quark charge matrix, $\lambda_4=\frac{1}{\sqrt 3}(\sqrt {2} \lambda_0 +\lambda_8)$
where $\lambda_i$ are Gell-Mann matrices and $\lambda_0 =\sqrt {2/3}$ diag(1,1,1), $g_{\rho}$=5.95
is the $\rho \to 2\pi$ coupling constant, $g_{\sigma} \approx 3$.

We will use the  matrix element of sub-process $\gamma^*(q,\mu) \to P(p)
\gamma(k,\nu)$
\ba
M^{\gamma^*\to P\gamma}=\frac{\alpha}{\pi f_{\pi}}F_P(q^2)
\varepsilon_{\mu\nu\alpha\beta}q^{\alpha}k^{\beta}\epsilon^\mu(k)\epsilon^\nu(q), \,\,f_{\pi}=93 MeV,
\ea
with condition $F(0)=1$. We remind the current algebra expression
for the pion decay width

$$\Gamma_{exp}^{\pi_0\to 2\gamma} \approx 7.3 \,\,eV.$$
NJL result is
$$\Gamma_{NJL}^{\pi_0\to 2\gamma}=\alpha^2M_\pi^3/(64\pi^3f_\pi^2)\approx 7.1 \,\,eV.$$

The similar expression for  $\gamma^*(q,\mu) \to S(p) \gamma(k)$:
\ba
M^{\gamma^*\to S\gamma}=\frac{\alpha}{\pi f_{\pi}}F_S(q^2)(g_{\mu\nu}\cdot k q - q_\nu \cdot k_{\mu})
\epsilon^\mu(k)\epsilon^\nu(q).
\ea
 Total cross sections of creation $P\gamma$, $S\gamma$ in
 electron-positron annihilation are:
 \ba
\sigma_{theor}^{e \bar e \to P\gamma}=\frac{8 \pi \alpha}{3 M_P^2}\Gamma_p^{\gamma\gamma}\left(1-\frac{M_P^2}{s}\right)^3 \frac{M_{\rho}^4}{(s-M_{\rho}^2)^2+M_{\rho}^2 \Gamma_{\rho}^2},
\ea
In the same may for scalar particles we obtain
\ba
\sigma_{theor}^{e \bar e \to S\gamma}=\frac{8 \pi \alpha}{3 M_S^2}\Gamma_S^{\gamma\gamma}\left(1-\frac{M_S^2}{s}\right)^3 \frac{M_{\omega}^4}{(s-M_{\omega}^2)^2+M_{\omega}^2 \Gamma_{\omega}^2},
\ea

The gauge invariant provide convergence of the loop momentum integral for $a_\mu$ so the application
of such a low-energy models as Nambu-Iona-Lasinio (NJL) one \cite{Volkov09} for
description of processes of conversion of a virtual photon to light mesons and in particular to
mesons and a real photons, NJL permits to calculate constant of strong coupling $g_{\pi}, g_{\rho}, g_{\sigma}$.

Calculations leads to
$$(g-2)_{\mu}^{\pi_0\gamma} \approx 4.5 \cdot 10^{-10},$$
$$(g-2)_{\mu}^{\eta\gamma} \approx 0.7 \cdot 10^{-10},$$
$$(g-2)_{\mu}^{\sigma\gamma} \approx 0.15 \cdot 10^{-10}.$$

The contribution out the experimentally accessed region $0.6<\sqrt{s}<1.03 \,\,GeV$
was obtained \cite{Akhmet}
\ba
a_\mu(\pi_0\gamma, 0.6<\sqrt{s}<1.03 GeV)^{exp}=(4.5\pm 0.15)\times 10^{-10}; \nn \\
a_\mu(\eta\gamma, 0.69<\sqrt{s}<1.33 GeV)^{exp}=(0.73\pm 0.03)\times 10^{-10}.
\ea

The contribution from the region below the experimentally accessible region is
\ba
a_\mu(\pi_0\gamma, \sqrt{s}<0.6 GeV)=(0.13 \pm 0.01)\times 10^{-10}.
\ea

The contribution of radiative processes with charged pions production and 2 neutral
ones was found \cite{Yndur02,Akhmet} to be is
\ba
a_\mu(e^+e^-\to \pi^+\pi^-\gamma, \sqrt {s} <1.2GeV)=(38.6\pm 1.0)\times 10^{-11}.
\ea

Note that $P\gamma$ we use only quark loops, whereas for $S\gamma$ besides quark loops the
triangle loops with pions and kaons as well are relevant.
Total contribution $\pi_0\gamma;\eta\gamma;2\pi\gamma$ is
\ba
a_{\mu}^{exp}(e^+e^-\to hadr +\gamma)=(93 \pm 1.0)\times 10^{-11}.
\ea

For process $e^+e^-\to \sigma \gamma$ S. Narison had obtained \cite{Naris}
starting from QCD sum rules, two different results one of the is:
\ba
a_\mu(\sigma(600)\gamma)=0.1 \cdot 10^{-10},
\ea
which is in agreement with our NJL approach.

As for $\gamma^* \to\rho\to\sigma\gamma$ the quark loops as well as loops with $\pi_\pm$,
$K_\pm$ must be taken into account, and, besides the imaginary part of meson loops amplitudes
must be taken into account, whereas for quark loops only real part mast be considered
(naive confinement). Both component of $\sigma$ meson $\sigma=\sigma_u\cos\alpha +
\sigma_s\sin\alpha $ contribute, besides $\sigma_s$ do not contain quarks and pion loops.
Main contribution arise from $\sigma_u.$
For the case $\sigma\gamma$ main contributions arises from light quarks
and from the pion loop with constructive interference, resulting
$\Gamma_{\sigma\to 2\gamma}=4.3 \,\,KeV$ \cite{Volkov09, Echaya}.

In NJL approach we obtain for $e^+e^- \to \sigma(550)\gamma$:
\ba
a_{\mu}(\sigma(550))=0.16 \times 10^{-10}.
\ea

We use the $\sigma$-meson mass $m_{\sigma}=550 \,MeV$ as well calculated in \cite{550}
and agreement with experiment \cite{Cine}.
Calculating $\gamma^* \to \pi^+\pi^- \gamma$ we use Born approximation and the experimental
pion form-factor \cite{amb}
\ba
(g-2)_{\mu}^{\pi^+\pi^- \gamma}=\frac{1}{4\pi^3}\int\limits_{4m_{\pi}^2}^{\infty}\sigma^{\pi^+\pi^-\gamma}(s) K(s)ds.
\ea
We use here \cite{sch,drees}
\ba
\sigma^{e^+e^- \to \pi^+\pi^- \gamma}(s) =\frac{2\alpha}{\pi}\sigma_{B}(s) \cdot \Delta(s); \nn \\
\sigma_B(s)=\frac{\pi \alpha^2 \beta^3}{3s}|F_{\pi}(s)|^2; \,\,\beta=\sqrt {1-\frac{4m^2}{s}}; \nn \\
\Delta=\frac{3}{4\beta^2}(1+\beta^2)-2\ln\beta +3\ln\frac{1+\beta}{2}+ \nn \\ \frac{1}{8\beta^3}(1-\beta)(-3-3\beta+7\beta^2-5\beta^3)L_{\beta}+\frac{1+\beta^2}{2\beta}F(\beta); \nn \\
F(\beta)=-2Li(\beta)+2Li(-\beta)-2Li(1+\beta)+2Li(1-\beta)+ \nn \\
3Li (\frac{1+\beta}{2})-3Li(\frac{1-\beta}{2})+3\xi_2, \,\,
\xi_2=\frac{\pi^2}{6}.
\ea
As a result, with $|F_{\pi}|^2=1,$ we obtain $(g-2)_{\mu}^{\pi^+\pi^-\gamma}=0.7 \times 10^{-10}$,
but with real form-factor \cite{amb}. $(g-2)_{\mu}^{\pi^+\pi^-\gamma}=3.13 \times 10^{-10},$ agreement
with contribution of nonresonance channel
\cite{Akhmet}.

Analog of semi hadronic contributions is the light by light (L-b-L)
scattering mechanism with intermediate states with scalar and pseudo
scalar mesons (Fig2).

Convergence of different recent model calculations lead to the result \cite{Bartosh}
(see \cite{China09}, A. Nyfeller talk and references there in)
\ba
a_\mu^{L-b-L}=(10.5\pm 2.6)\times 10^{-10}.
\ea
We put below the definite contributions (we follow the paper \cite{Bartosh}):
\ba
\pi_0: 81.8\times 10^{-11}; \eta: 5.62\times 10^{-11}; \eta':(8\pm 1.7)\times 10^{-11}; \nn \\
\sigma(600): 11.67\times 10^{-11}; a_0(980): 0.62\times 10^{-11},
\ea
with the total contribution
\ba
a_\mu^{LbL}=(107.74 \pm 16.81)\times 10^{-11}.
\ea

\section{Acknowledgements}
We are grateful for the supported in part by the grant from RFFI (Grant No.10-02-00717).

\begin{figure}
\includegraphics[width=0.5\textwidth]{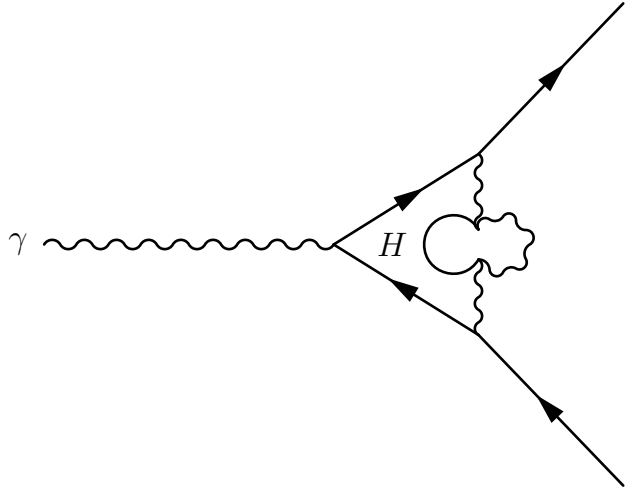}
\caption{Contributions from state  $\gamma^* \to P;S; \pi^+\pi^-; \gamma.$ \qquad \\
Where $H=\pi^0; \eta; \pi^+\pi^-.$}
\label{Fig1}
\end{figure}

\begin{figure}
\includegraphics[width=0.8\textwidth]{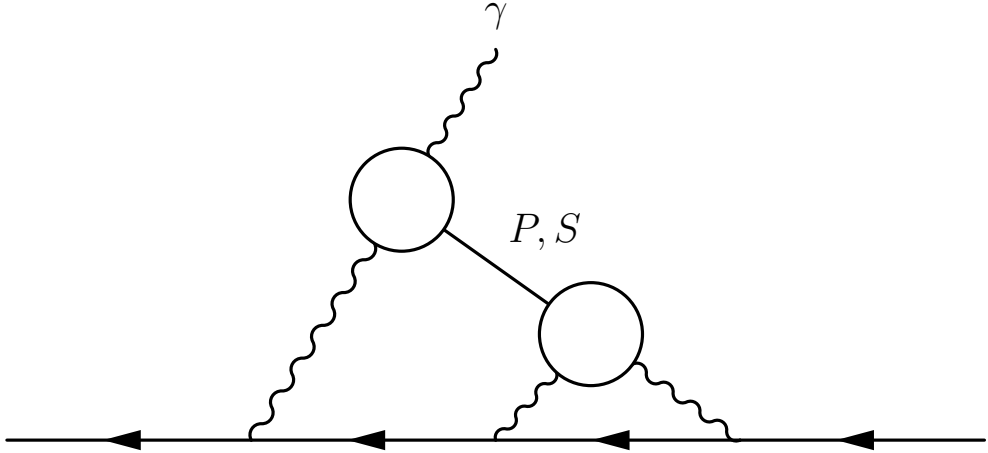}
\caption{Contributions type L-b-L mechanism with intermediate state $P; S$}
\label{Fig2}
\end{figure}


\begin{thebibliography}{10}
\bibitem{Brodsky}
S.~J.~Brodsky, E.~de
~Rafael, Phys. Rev. 168 (1968)1620
\bibitem{Achasov02}
N.~Achasov and A.~Kiselev,
Phys.Rev. D 65,097302 (2002).
\bibitem{Naris}
S.~Narison, Phys. Lett. B 568(2003),231-236.
\bibitem{Yndur02}
F.~Troconiz and F.~Yndurain, Phys.Rev. D 65,093001, (2002).
\bibitem{Akhmet}
R.~Akhmetishin et. al, Phys. Lett. B 605(2005),26-36.
\bibitem{IMP}
Yu.M.Bystritskiy, E.A.Kuraev, M.Secansky, M.K.Volkov,
Int.J.of Modern Physics A 24 (2009)2629.
\bibitem{CEPhys}
M.K.Volkov, E.A.Kuraev, Yu.M.Bystritskiy,
Cent.Eur.J.Phys., \,\,\,arXiv: 0904.2484 [hep-ph].
\bibitem{Volkov09}
M.~Volkov et. al, \,\,\,arXiv: 0901.1981 [hep-ph].
\bibitem{Echaya}
M.K.Volkov, Fiz. Elem. Chast. Atom. Yadra, 17 (1986)433
\bibitem{Prog}
D.Ebert, H.Reinhardt, M.K.Volkov,
Prog. Part. Nucl. Phys., Vol.33, (1994)1-120.
\bibitem{UFN}
M.K.Volkov, A.E.Radzhabov,
Phys. Usp. 49 (2006) 551
\bibitem{amm04}
A.Nyffeler, Nucl.Phys. Proc.Suppl. {\bf 131},162 (2004)
\bibitem{Bartosh}
E.~Bartosh,et. al,Nucl. phys. B632 (2002),330.
\bibitem{China09}
International Workshop on $e^+e^-$ collisions from $\Phi$ to $\Psi$, 2009.
\bibitem{550}
M.K.Volkov, M.Nagy, V.L.Yudichev, Nuovo Cim. A112 (1999) 225.
\bibitem{Cine}
Hua-Xing Chen, Atsushi Hosaka, Hiroshi Toki, Shi-Lin Zhu, \,\,\,
arXiv: 0912.5138 [hep-ph].
\bibitem{sch}
Schwinger J. particles, Sources, And Fields, 1998, v2, Westview Press.
\bibitem{drees}
Drees M, and Hikasa K., Phys.Lett. B252 (1990)127.
\bibitem{amb}
F.Ambrosino et al., Phys.Lett B670 (2009)285.

\end{thebibliography}
\end{document}